\PassOptionsToPackage{pdfpagelabels=false}{hyperref}
\documentclass[fleqn,usenatbib]{mnras}
\usepackage{newtxtext,newtxmath}
\usepackage{setspace}

\usepackage[T1]{fontenc}
\usepackage{ae,aecompl}
\usepackage{tabularx}
\usepackage{multicol}

\usepackage{graphicx}	
\usepackage{amsmath}	
\usepackage{amssymb}	

\newcommand*{\mhalo}{\ensuremath{M_\text{h}}}
\newcommand*{\msol}{\text{M}\ensuremath{_\odot}}
\newcommand*{\hinv}{\ensuremath{h^{-1}}}
\newcommand*{\hmsol}{\hinv\,\msol}
\newcommand*{\hmpc}{\hinv\,\text{Mpc}}
\newcommand*{\vpeak}{\ensuremath{V_\text{peak}}}
\newcommand*{\mgal}{\ensuremath{M_\ast}}
\newcommand*{\slog}[2]{\ensuremath{\sigma[#1|#2]}}
\newcommand*{\slogvm}{\slog{\vpeak}{\mgal}}
\newcommand*{\slogmv}{\slog{\mgal}{\vpeak}}
\newcommand*{\slogmm}{\slog{\mgal}{\mhalo}}
\newcommand*{\fsat}{\ensuremath{f_\text{sat}}}
\newcommand*{\kms}{\text{km}\,\text{s}\ensuremath{^{-1}}}
\newcommand*{\mybox}{\framebox}

\title[Galaxy-halo scatter at Milky Way masses]{Constraining the scatter in the galaxy-halo connection at Milky Way masses}

\author[Cao et al]
{
Jun-zhi Cao,$^{1}$
Jeremy L. Tinker,$^{1}$
Yao-Yuan Mao,$^{2,3,4}$\thanks{NHFP Einstein Fellow}
and Risa H. Wechsler$^{5,6,7}$
\vspace*{6pt}
\\
$^{1}$Center for Cosmology and Particle Physics, Department of Physics, New York University, New York, NY, USA \\
$^{2}$Department of Physics and Astronomy, Rutgers, The State University of New Jersey, Piscataway, NJ 08854, USA\\
$^{3}$Department of Physics and Astronomy, University of Pittsburgh, Pittsburgh, PA 15260, USA\\
$^{4}$Pittsburgh Particle Physics, Astrophysics, and Cosmology Center (PITT PACC), University of Pittsburgh, Pittsburgh, PA 15260, USA\\
$^{5}$ Department of Physics, Stanford University, 382 Via Pueblo Mall, Stanford, CA 94305, USA \\
$^{6}$ Kavli Institute for Particle Astrophysics and Cosmology, Stanford University, Stanford, CA 94305, USA \\
$^{7}$ SLAC National Accelerator Laboratory, Menlo Park, CA 94025, USA
}

\date{\today}

\pubyear{2019}

\begin{document}

\label{firstpage}
\pagerange{\pageref{firstpage}--\pageref{lastpage}}
\maketitle
\begin{abstract}

We develop and implement two new methods for constraining the scatter in the relationship between galaxies and dark matter halos. These new techniques are sensitive to the scatter at low halo masses, making them complementary to previous constraints that are dependent on clustering amplitudes or rich galaxy groups, both of which are only sensitive to more massive halos. In both of our methods, we use a galaxy group finder to locate central galaxies in the SDSS main galaxy sample. Our first technique uses the small-scale cross-correlation of central galaxies with all lower mass galaxies. This quantity is sensitive to the satellite fraction of low-mass galaxies, which is in turn driven by the scatter between halos and galaxies. The second technique uses the kurtosis of the distribution of line-of-sight velocities between central galaxies and neighboring galaxies. This quantity is sensitive to the distribution of halo masses that contain the central galaxies at fixed stellar mass. Theoretical models are constructed using peak halo circular velocity, $\vpeak$, as our property to connect galaxies to halos, and all comparisons between theory and observation are made after first passing the model through the group-finding algorithm. We parameterize scatter as a log-normal distribution in $\mgal$ at fixed $\vpeak$, $\slogmv$.
The cross-correlation technique yields a constraint of $\sigma[ \mgal|\vpeak]=0.27\pm 0.05$ dex at a mean $\vpeak$ of $168$\,\kms, corresponding to a scatter in $\log \mgal$ at fixed $\mhalo$ of $\sigma[ \mgal|M_{\rm h}]=0.38\pm 0.06$ dex at $\mhalo=10^{11.8}\,\msol$. The kurtosis technique yields $\sigma[ \mgal|\vpeak]=0.30\pm0.03$ at $\vpeak=209$\,\kms, corresponding to $\sigma[ \mgal|M_{\rm h}]=0.34\pm 0.04$ at $\mhalo=10^{12.2}\,\msol$. The values of $\sigma[ \mgal|M_{\rm h}]$ are significantly larger than the constraints at higher masses, in agreement with the results of hydrodynamic simulations. This increase is only partly due to the scatter between $\vpeak$ and $M_{\rm h}$, and it represents an increase of nearly a factor of two relative to the values inferred from clustering and group studies at high masses.
\end{abstract}

\section{Introduction}\label{sec:Introduction}

The galaxy--halo connection plays an important role in inferring both galaxy formation physics and the underlying cosmological model from observations of galaxies and their distribution. Hitherto, galaxy redshift surveys have provided extensive data on the properties and spatial distribution of galaxies (e.g., \citealt{york2000sloan}, \citealt{dawson2012baryon}, \citealt{dawson2016sdss}).  Understanding these data requires a model for the relationship between galaxies and the unseen dark matter. In a recent review, \cite{wechsler2018connection} laid out the various methods to predict or constrain this galaxy--halo connection. These include, in order of decreasing physical complexity: hydrodynamic simulations, semi-analytic models, empirical forward models, sub-halo abundance matching, and halo occupation models. The first two methods attempt to model galaxy formation from first principles, while the last two intentionally remove as much physics as possible, using data-driven models to reverse-engineer the galaxy formation problem. These latter approaches are complementary to the first two methods, making predictions that can be tested by the more physical models in ways that are not straightforward using direct comparison between models and the data. In this paper, we present new measurements of the scatter in the galaxy--halo connection. At present, physical and empirical models are in good agreement for the mean of the stellar-to-halo mass relation (SHMR) as a function of halo mass. However, the scatter about this relation is not well constrained. Observational constraints on the scatter in $\log \mgal$ at fixed halo mass, 
$\sigma[\mgal|\mhalo]$, are relegated to samples of galaxies at high halo and galaxy masses. These studies have yielded values of $\sigma[\mgal|\mhalo]\lesssim 0.2$ (\mbox{\citealt{reddick2013connection}}, \mbox{\citealt{zu2015mapping}}, \mbox{\citealt{tinker2017correlation}},  \mbox{\citealt{lange2018updated}}). To date, these results reflect the galaxy--halo connection in a regime dominated by massive, red quenched galaxies.

In this work, we present two new methods to constrain scatter at low galaxy and halo masses, $\mhalo\lesssim 10^{12}$ $\hmsol$. In this regime, the galaxy population is dominated by star-forming objects. Thus, the scatter at these masses are more reflective of the physics of star formation, and how these processes are correlated with the growth of the dark halo. The more correlated $\dot{\mgal}$ is to $\dot{\mhalo}$, the smaller the scatter between the two. 

Here, we use subhalo abundance matching to model the galaxy--halo connection. Abundance matching assumes a relationship between galaxies and halos based on their rank-ordering: the most massive galaxies live in the biggest halos, but with some scatter between the the halo and galaxy properties used. For this approach, the choice of which halo property to match to $\mgal$ is critical. \cite{reddick2013connection} have shown that abundance matching models constructed by using various definitions of halo mass do not reproduce observational clustering results, even when assuming zero scatter (see also \citealt{zentner2016constraints} and \citealt{lehmann2016concentration}). Among individual halo properties, the peak value of a halo's circular velocity over its formation history, $\vpeak$, yields the best match to observations. However, one can create a single proxy that is a combination of multiple halo properties that also matches observed clustering (\citealt{lehmann2016concentration}). Here we limit our study to single halo properties, with $\vpeak$ as our proxy for $\mgal$. We quantify the scatter as  $\sigma[\mgal|\vpeak]$. In this paper, we will refer to a number of different scatter values between various properties. All scatter distributions are log-normal, but for brevity we will reference them as $\slog{X}{Y}$, or ``the scatter in $\log X$ at fixed $Y$.'' When constructing abundance matching model, we incorporate scatter as $\sigma(\log \vpeak)$, which represents constant scatter in $\log \vpeak$ of all halos. We present our method for the abundance matching model that yields $\sigma[\mgal|\vpeak]$ a constant for all halos in Appendix A.

This paper is organized as follows: The first section is the data section \ref{sec:Data}, which include the data we are using, the stellar mass function in abundance matching \ref{sec:Stellar} and the group finder algorithm (to detect central and satellite galaxies) \ref{sec:Galaxy}. The second section \ref{sec:Theoretical} is the theoretical models we adopt, including N-body simulations\ref{sec:N-body} in our mocks, the abundance matching method \ref{sec:Abundance} and the scatter in galaxy-halo connection \ref{sec:Stellar-to-halo}. Then we will move to the next section. This section include our two techniques to constrain scatter \ref{sec:Observables}. The first technique is using the ratio between two different measurements of $w_p$ \ref{sec:ratio}. The second technique is to use the kurtosis, $\kappa$, of line-of-sight velocity of neighbor galaxies near central galaxy \ref{sec:Line-of-sight}. Here kurtosis is the Fisher kurtosis, which means kurtosis for a normal distribution is zero. After that, we include our results for the two techniques, and compare them to results in other works \ref{sec:results}. Finally, we discuss the meaning of our work, as well as pros and cons of our techniques. In analysis of observational data we assume $\Omega_m=0.3$ and $h=0.7$ for distances and stellar masses.

\section{Data}\label{sec:Data}

\subsection{NYU-VAGC catalog and stellar mass}\label{sec:NYU-VAGC}

We use the spectroscopic data in SDSS DR7 (\citealt{abazajian2009seventh}), as well as the NYU-VAGC catalog in \cite{blanton2005new}. From the data, we construct a volume limited sample that includes all galaxies with $M_r-5\log h<-17.48$, yielding a redshift range of 0.01 to 0.033. We then further restrict the volume-limited sample to be complete in stellar mass as well, with the lower bound of the stellar mass is $\log \mgal=9.7$. Stellar masses are derived from the Principal Component Analysis (PCA) method of \cite{chen2012evolution}. \cite{tinker2017correlation}  compared all 14 different stellar masses available in the standard SDSS-III pipeline. They found that galaxy samples defined by the PCA masses had the highest clustering amplitude at fixed number density, implying that the PCA masses have the lowest observational scatter of all the available stellar masses, motivating the choice for our paper.

After construction of the stellar mass-limited sample, two subsamples are defined: `central' galaxies and `tracer' galaxies. Central galaxies are defined as centrals by their classification from a galaxy group finder, which we will describe presently. These central galaxies have $\log \mgal$ between 10.35 and 10.65.  Tracer galaxies are all galaxies with $\log \mgal$ lower than 10.35 but higher than 9.7. We focus on central galaxies in this mass range because they are expected to live in halos with $\mhalo \sim 10^{12}$ $\msol$.

\subsection{Stellar mass function}\label{sec:Stellar}

Abundance matching requires a measurement of the abundance of galaxies as a function of their stellar mass.
To estimate the stellar mass function (SMF), we use the $1/{\mathcal V}_{\rm max}$ method, where ${\mathcal V}_{\rm max}$ is the maximum volume to which a given target could be observed in SDSS. In each bin in $\log [\mgal]$, the mass function is defined as
\begin{equation}
\label{eq:smf}
\Phi_i (\log \mgal) \, \Delta \log \mgal= \Sigma_j \left[ \frac{1}{{\mathcal V}_{\rm max,j}} \right], 
\end{equation}

\noindent where $j$ represents all galaxies in bin $i$. To evaluate Eq.~\eqref{eq:smf}, we use the $V_{\rm max}$ values supplied in the VAGC. Although the SDSS survey is not complete for low surface brightness galaxies (\citealt{blanton2005new}), this does not influence our estimation for the stellar mass function given our sample limits.

We estimate the errors using the jackknife sampling technique. We divide the SDSS footprint into $5 \times 5$ nearly equal subsamples, 5 in RA and 5 in Dec. We find that the jackknife errors are consistent with Poisson at $\log \mgal >11.5$, but are about three times higher at lower stellar masses. We use the jackknife subsamples to estimate the full covariance matrix of the stellar mass function.

\subsection{Galaxy group finder}\label{sec:Galaxy}

Our methods of constraining scatter require knowledge of which galaxies are central galaxies, located at the center of their host halos. To separate central and satellite galaxies in the SDSS data, we use the galaxy group finder algorithm in \cite{tinker2011halo}, based on the method of \cite{yang05}. This group finder has been thoroughly tested and shown to yield accurate values for the satellite fraction of galaxies (\citealt{campbell2015assessing}). The group finder returns a value of $P_{\rm sat}$,  the probability a galaxy is a satellite galaxy. Standard results use a break point of $P_{\rm sat}=0.5$ to separate all galaxies into central and satellite galaxies. However, this method leads to around 15 percent impurities in the sample. For our fiducial results in this paper, we define our sample of central galaxies as those with $P_{\rm sat}<0.01$, which yields a high-purity sample with only a limited reduction in completeness (\citealt{tinker2017halo}). Full statistics of the galaxy number counts in our volume-limited sample are given in Table~\ref{tab:number_counts}. 

Even with our restricted definition of central galaxies, there will be some intrinsic bias in detecting the real central and satellite galaxies. For all comparisons between theory and observations, mock galaxies are passed through the group finder first, and the same criteria are used to define the sample of mock central galaxies and mock tracer galaxies. 

\begin{table}
\centering
\renewcommand{\arraystretch}{1.2}
\resizebox{\columnwidth}{!}
{\begin{tabular}{rlrrrr}
 \hline
Name & $\log \mgal$ & All & $P_{\rm sat}<0.5$ & $P_{\rm sat}<0.01$ & $P_{\rm sat}>0.5$  \\
 \hline
 \hline
 Full &  $>$9.7 & 16970 & 12344 & 10623& 4626 \\ 
  \mybox{Central} & 10.35--10.65 & 3573 & 2629 & \mybox{2250}& 944  \\
  \mybox{Tracer} & 9.7--10.35 & \mybox{9046} & 6254 & 5251& 2792 \\
 \hline
\end{tabular}}
\renewcommand{\arraystretch}{1}
\caption{Galaxy number counts in our volume-limited sample. $P_{\rm sat}$ is the probability of a galaxy is a satellite galaxy, returned from the group finder. Galaxies with $P_{\rm sat}>0.5$ are categorized as satellites. Here we use galaxies with $P_{\rm sat}$ smaller than 0.01 to be our central galaxy sample to avoid impurities. The numbers in the two samples used here are indicated by the boxes.} 
\label{tab:number_counts}
\end{table} 

\section{Theoretical Models}\label{sec:Theoretical}

We construct theoretical models on the abundance matching framework, matching $\vpeak$ to the stellar mass function described in the previous section. In this section we will describe the multiple simulations that we use, as well as a modified method of implementing abundance matching. We pay special attention to the relationships between different forms scatter, depending on whether one is binning on halo properties or galaxy properties, and whether one uses halo mass or $\vpeak$.

\subsection{\texorpdfstring{$N$}{N}-body simulations}\label{sec:N-body}

We use three simulations to construct our mock catalogs, listed in Table 2. The Bolshoi-Planck (hereafter BolshoiP) simulation is a high-resolution simulation with a box size of $(250\,\hmpc)^3$ and a density field resolved with $2048^3$ particles (\citealt{klypin2016multidark}). C250, described in \cite{lehmann2016concentration}, has the same box size as BolshoiP, but with 2560$^3$ particles, yielding a factor of two higher mass resolution. C125 has the same number of particles as the BolshoiP, and one-eighth of the volume (hence a factor of 8 higher resolution) . While this smaller volume limits its utility for comparison to the data, C125 shows that the predictions from BolshoiP and C250 are not biased due to resolution limits. 

Halos are detected by using the {\sc Rockstar} algorithm (\citealt{behroozi2012rockstar}) and merger histories are constructed with {\sc Consistent-Trees} (\citealt{behroozi2012gravitationally}). These merger trees are used to determine the $\vpeak$ value of each halo and subhalo. The boundary of a halo is defined by the $\Delta = 200 \rho_{\rm crit}$ standard.

\begin{table*}
\begin{center}
\renewcommand{\arraystretch}{1.2}
\begin{tabular}{rlrrrrrrr}
 \hline
Name & $\Omega_m$  & $\Omega_b$ & $\sigma_8$ & $h$ & $L_{\rm box}\;[{\rm Mpc} h^{-1}]$ & $N_{\rm p}$ & $m_p\;[\hmsol]$ \\
 \hline
 \hline
 Bolshoi-Planck &  0.295  & 0.048 & 0.823& .678 & 250 & 2048$^3$ &  1.49$\times 10^8$ \\ 
  C250 & 0.295  & 0.047 & 0.834& .688 & 250& 2560$^3$& 7.63$\times 10^7$ \\
  C125 & 0.286  & 0.047 & 0.82& .7 & 125& 2048$^3$& 1.8$\times 10^7$ \\
 \hline
\end{tabular}
\renewcommand{\arraystretch}{1}
\caption{Cosmological and simulation parameters for the simulations used in the analyses.  All models are flat $\Lambda$CDM cosmologies.} 
\end{center}
\end{table*}

\subsection{Abundance matching and the incorporation of scatter}\label{sec:Abundance}

Abundance matching, in its simplest form, equates the cumulative stellar mass function and cumulative halo property function, then uses this result to assign stellar mass to each halo in the simulation, i.e., 
\begin{equation}
\label{e.am}
\int_{\vpeak}^{\infty}n(\vpeak')d\vpeak' = \int_{\mgal}^{\infty}\Phi (\mgal')d\mgal'.
\end{equation}

\noindent Here we use $\vpeak$ as the halo property, but any halo property can be used in principle. However, Eq.~\eqref{e.am} assumes no scatter between $\vpeak$ and $\mgal$. Typically, scatter is modeled as a distribution of stellar mass at fixed $\vpeak$. Parameterizing the scatter as a log-normal distribution of $\mgal$ at fixed halo property has proven successful in modeling galaxy clustering measurements, and it is consistent with modern hydro-dynamic simulations (\citealt{wechsler2018connection}). 
To implement such a model, one would need to add a scatter value, sampled from the log-normal distribution, to $\mgal$ value assigned to each halo given Eq.~\eqref{e.am}. However, this procedure will alter the overall stellar mass function because it is equivalent to convolving stellar mass function with a log-normal function. In order to preserve the stellar mass function, one can first deconvolve the stellar mass function and then use the deconvolve the stellar mass function in Eq.~\eqref{e.am}, such that after adding scatter the overall stellar mass function would match the original \citep[e.g.,][]{behroozi2010comprehensive}.

In this work we adopt a different procedure that is more convenient operationally: we directly add scatter to the halo property before the matching procedure. Since the matching procedure is done in the last step, we ensure that the input stellar mass function is always preserved. However, the scatter value that we add to the halo property (in our case, $\vpeak$) is not equal to the desired scatter $\sigma[\mgal|\vpeak]$ (as they are scatter of different quantities). Thus, we have developed a technique to empirically map $\sigma[\mgal|\vpeak]$ to $\sigma[\log \vpeak]$, as a function of $\vpeak$. This technique allows us to obtain the desired $\sigma[\mgal|\vpeak]$ and preserve the stellar mass function as the same time. A detailed description of the technique is given in Appendix A.

\begin{figure*}
\includegraphics[width= \textwidth]{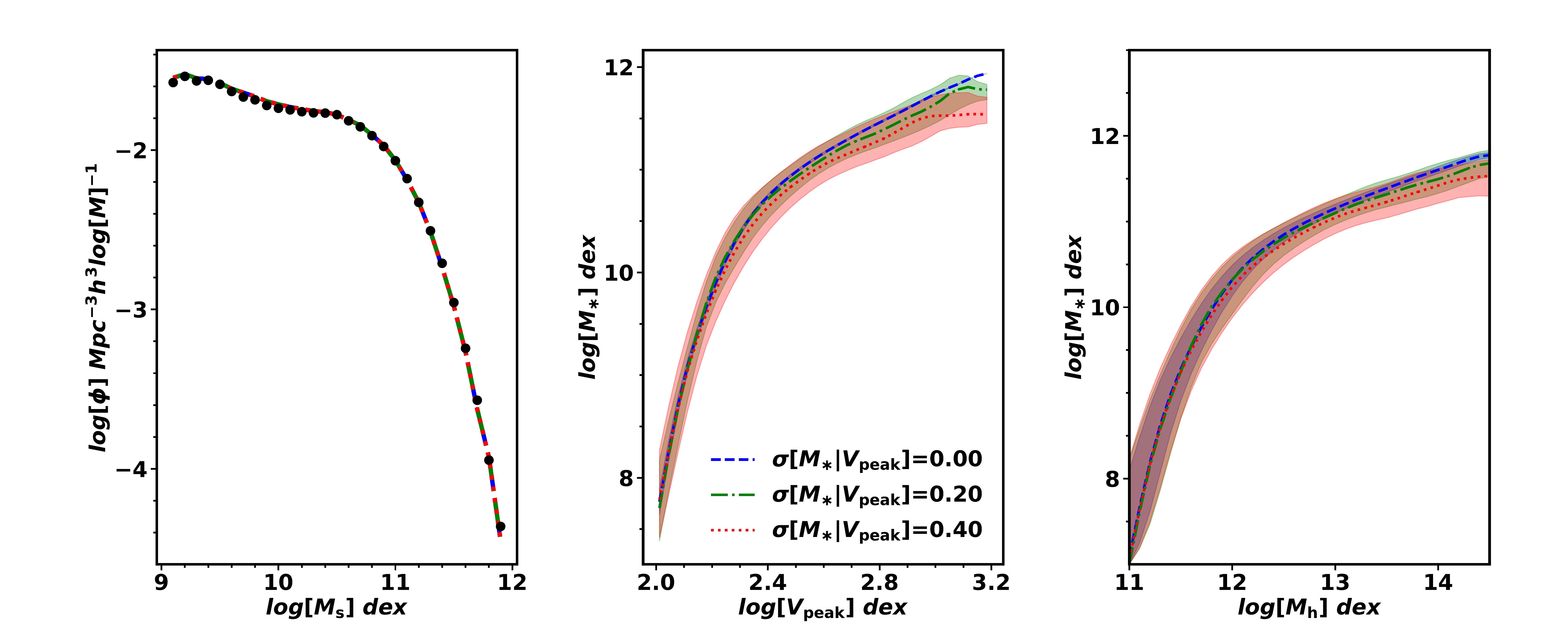}
\caption{Stellar mass function and SHMR for models with different values of scatter. {\em Left panel:} Stellar mass function from our model with $\slogmv$ values of 0, 0.2, and 0.4 dex. Black dots are the observed SDSS PCA stellar mass function and curves are the abundance matching models. {\em Middle panel:}  Relation between stellar mass and $\vpeak$. The dots in the middle row are averaged relation between $\mgal$ and $\vpeak$. Shaded areas show $\sigma[ \mgal|\vpeak]$ at different $\vpeak$. There is no shaded area in the first panel of the middle row since $\mgal$ are from abundance matching using $\vpeak$ without scatter. The bottom panel, shows the relation between stellar mass and $M_{\rm h}$. Due to the scatter between $\vpeak$ and $M_{\rm h}$, the behaviors of the shaded areas in the right panel and the middle panel are different. As for the blue line and shaded area in the right panel, there is a significant scatter between $M_{\rm h}$ and $\mgal$ at halo mass smaller than $10^{13} \msol$ even $\sigma[ \mgal|\vpeak]=0$. And $\sigma[ \mgal|M_{\rm h}]=0$ will go to zero as halo mass goes beyond $10^{13} \msol$}
\label{Mh_Ms_SMF_paper_v1}
\end{figure*}

\begin{figure*}
\includegraphics[width= \textwidth]{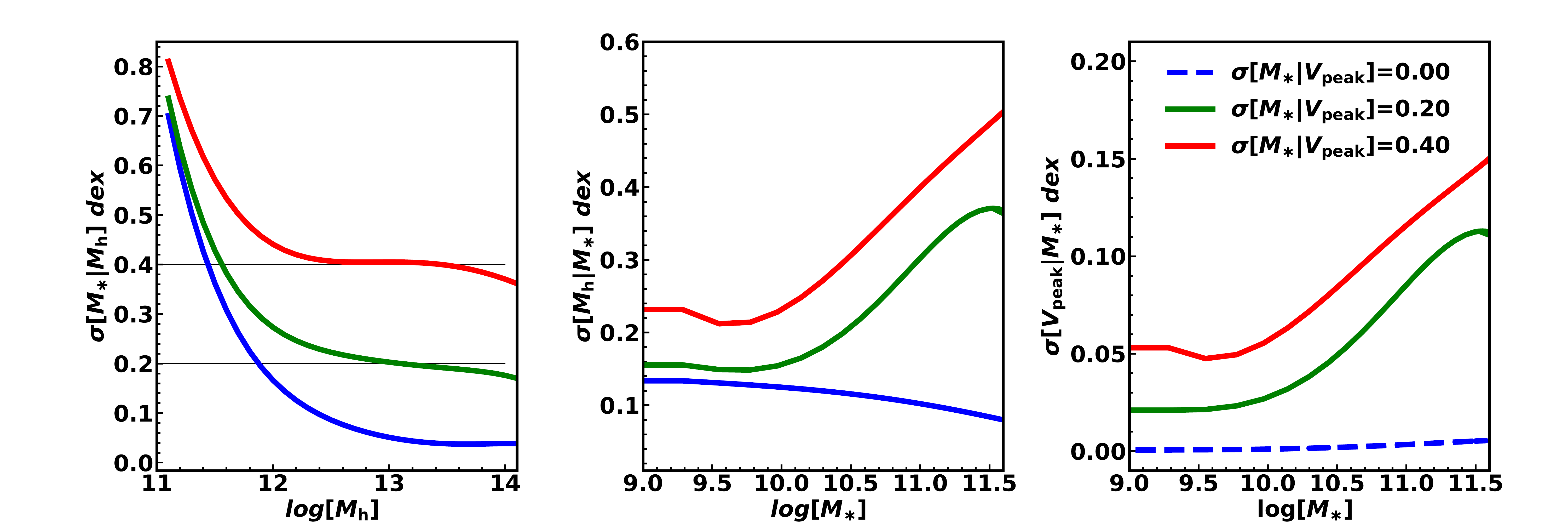}
\caption{Left-hand panel: $\sigma[ \mgal|M_{\rm h}]$ as a function of $M_{\rm h}$. Lines with different colors are with different $\sigma[\mgal|\vpeak]$ values. $\sigma[ \mgal|M_{\rm h}]$ goes down as a function of $M_{\rm h}$, and gets close to $\sigma[\mgal|\vpeak]$ at large halo mass. Middle panel: Similar to the left panel, but the $x$-axis is $\mgal$. The blue line ($\sigma[\mgal|\vpeak]=0$) is decreasing with $\mhalo$ due to intrinsic scatter between halo mass and $\vpeak$. Right-hand panel:  The relation between $\sigma[ \vpeak|\mgal]$ and $\mgal$ at fixed $\sigma[ \mgal|\vpeak]$. The green and purple lines are increasing as a function of $\mgal$ due to the slope changes in the mean relation between $\mgal$ and $\vpeak$. The blue line is $\sigma[ \vpeak|\mgal]$ as a function of $\mgal$ for $\sigma[\mgal|\vpeak]=0$, and it should always be zero.}
\label{scatter_without_data_v1}
\end{figure*}



\subsection{Stellar-to-halo mass relation}\label{sec:Stellar-to-halo}

We start by applying the abundance matching technique to the halo catalogs with no scatter between $\mgal$ and $\vpeak$. These results are shown in the left-hand column of panels in Figure \ref{Mh_Ms_SMF_paper_v1}. The top panel shows our measurement of the stellar mass function, as well as the abundance matching fit to these data. The incompleteness of SDSS data is apparent at the very lowest mass scales, but in the range $\log \mgal=[9.5,10]$, there is a clear upturn in the SMF, consistent with that measured in the luminosity function (\citealt{blanton2005new}) as well as in other SMF measurements (\citealt{drory2009bimodal}, \citealt{moustakas2013primus}, \citealt{baldry2008galaxy}). The SHMR, whether plotted as a function of $\mhalo$ of $\vpeak$, shows the expected behavior or a steep slope as low masses, and a break to a shallower slope at high masses, with this pivot point occurring at $\mhalo\sim 10^{12}$ $\hmsol$. There is scatter between $\mhalo$ and $\vpeak$, as indicated by the shaded region in the SHMR. The other columns in Figure \ref{Mh_Ms_SMF_paper_v1} show models with $\sigma[\mgal|\vpeak]=0.2$ and $0.4$. For each model, the shape of the stellar mass function is identical to that of the zero-scatter model, demonstrating the efficacy of our approach. 

Figure \ref{scatter_without_data_v1} shows different quantities that measure the scatter for models with $\sigma[\mgal|\vpeak]=0$, 0.2, and 0.4. At high halo masses, the scatter in $\mgal$ at fixed $\mhalo$ is the same as at fixed $\vpeak$, reflecting the fact that the scatter between these two halo properties monotonically decreases with increasing halo mass. Below $\mhalo=10^{12}$ $\msol$, however, $\sigma[\mgal|\mhalo]$ increases rapidly. Other panels in Figure \ref{scatter_without_data_v1} show $\sigma[\mhalo|\mgal]$ as well as $\sigma[\vpeak|\mgal]$ as a function of $\mgal$. We note that the small values of $\sigma[\vpeak|\mgal]$ are driven by the fact that $\vpeak\sim \mhalo^{0.35}$, thus the scatter in $\log\vpeak$ is roughly 1/3 the scatter in $\log \mhalo$.

\section{Observables for Constraining the Scatter}\label{sec:Observables}

We present two observables that are sensitive to the scatter in the galaxy-halo connection at the low-mass end. (1) The ratio of the auto correlation of central galaxies with the cross correlation of those central galaxies with the tracer galaxies, and (2) the kurtosis of the pairwise velocity distribution around central galaxies. One quantity is projected while the other is fully line-of-sight, making the combination complementary. We describe those two approaches in the following subsections.

\subsection{\texorpdfstring{$w_p$}{wp} ratio at small scales}\label{sec:ratio}

For the first technique to constrain scatter we utilize the two-point correlation function projected along the line of sight, $w_p(r_p)$. We use the ratio between two different measurements of $w_p$, and we refer to this ratio as $\omega$, defined as the ratio between the auto correlation of the central galaxy sample ($w_p^{\rm auto}$), and the cross correlation between central galaxies and tracer galaxies ($w_p^{\rm cross}$) 

\begin{equation} \label{eq:omega}
 \omega = \frac{w_p^{\rm auto}({r_p \leqslant 0.6 \, \text{Mpc}\, h^{-1}})}{w_p^{\rm cross}({r_p \leqslant 0.3 \, \text{Mpc}\, h^{-1}})} 
\end{equation}

\noindent Note that both $w_p$ quantities in the ratio are a single bin with lower limit $r_p=0$. We use a different outer limit for the numerator and denominator to reduce the error bar for the $w_p^{\rm auto}$, which is impacted by the lack of pairs of central galaxies at small $r$. We note that $w_p^{\rm auto}$ is largely independent of $r_p$ at small scales\footnote{In the three-dimension real-space correlation function, there are no central galaxy pairs within $R_{\rm vir}$ of the halos. However, with $w_p(r_p)$ being a projected quantity, $w_p^{\rm auto}$ goes to a constant at small $r_p$. In the language of halo occupation, this is the `two-halo term.' See, for example, Figure 3 of \cite{zehavi2004} for a breakdown of the one and two-halo terms in $w_p$.}. The exact choices of bin size were set to construct a quantity that offers maximal constraining power on scatter. Specifically, we want a quantity that balances the dependence on scatter with the precision with which we can measure it. The former pushes us to smaller scales, but the latter requirement prevents the $r_p$ limit from being entirely in the shot-noise regime. Thus, the limits in Eq.~\eqref{eq:omega} put $\omega$ near the middle of the one-halo term.

\begin{figure*}
\includegraphics[width= \textwidth]{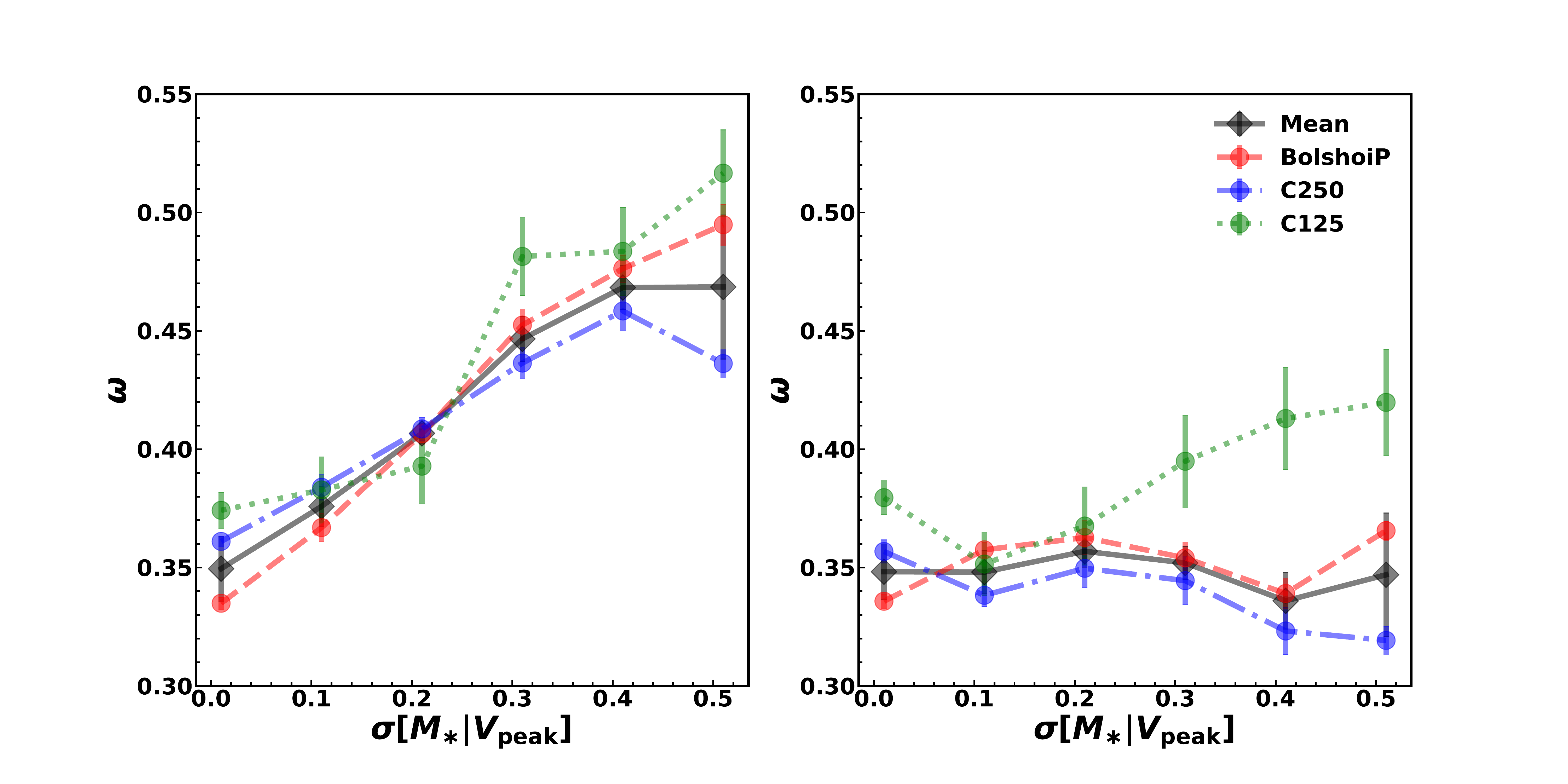}
\caption{The ratio of the projected two-point function from $V_{\rm{peak}}$ abundance matching as a function of scatter. The ratio $\omega$ is defined in Eq.~\eqref{eq:omega}. In the first panel, the ratio increases as the scatter increases. This is because ${\rm wp}_{\rm cross}$ decreases as the scatter increases, while ${\rm wp}_{\rm auto}$ is almost a constant. The black dots are the average of the values from the three simulations we used, weighted by the box size of the simulations. The second panel shows the results when only adding scatter to the central galaxies, with no scatter in the tracer sample. Figure \ref{high_mass_only_v1} shows two different ways to construct this test, both of which give consistent results. 
}
\label{wp_all_in_one_Vpeak_without_data_v1}
\end{figure*}

The left-hand panel of Figure \ref{wp_all_in_one_Vpeak_without_data_v1} shows how $\omega$, defined in Eq.~\eqref{eq:omega}, depends on scatter in our abundance matching models. There is a nearly linear dependence of $\omega$ on $\slogmv$. The sensitivity of $\omega$ to scatter is due to the sensitivity of the statistic to the satellite fraction of galaxies, $\fsat$; as scatter increases, $\fsat$ decreases (\citealt{reddick2013connection}). A smaller number of satellites means fewer central-satellite pairs in $w_p^{\rm cross}$, but the amplitude of $w_p^{\rm auto}$ remains unchanged. Using the ratio of correlation functions removes any dependence $w_p$ may have on cosmology or other properties that the overall scale of clustering is sensitive to. In this panel, we show the volume-weighted mean of all three simulations. We estimate the error bar by dividing the two larger simulations into 8 sub-volumes (each equivalent to the C125 simulation), and find the error in the mean of $(8+8+1)=17$ sub-volumes. We also show the individual results from BolshoiP, C250 and C125. Although C125 is significantly noisier than the large simulations, the general dependence of $\omega$ with $\slogmv$ is clear, and the deviations of the C125 results from the mean are consistent with statistical noise. Thus, resolution effects on the halo substructure in C250 and BolshoiP are not impacting our results.

As can be seen in the results, the differences between the C250 and BolshoiP results, while similar, are not consistent with statistical fluctuations; the differences are somewhat larger than the combined errors on both simulations. This difference may be due to the different cosmologies of the two simulations. Thus, when comparing model predictions to data, we use the difference between C250 and BolshoiP as the error in the prediction to be conservative.

\begin{figure*}
\includegraphics[width= \textwidth]{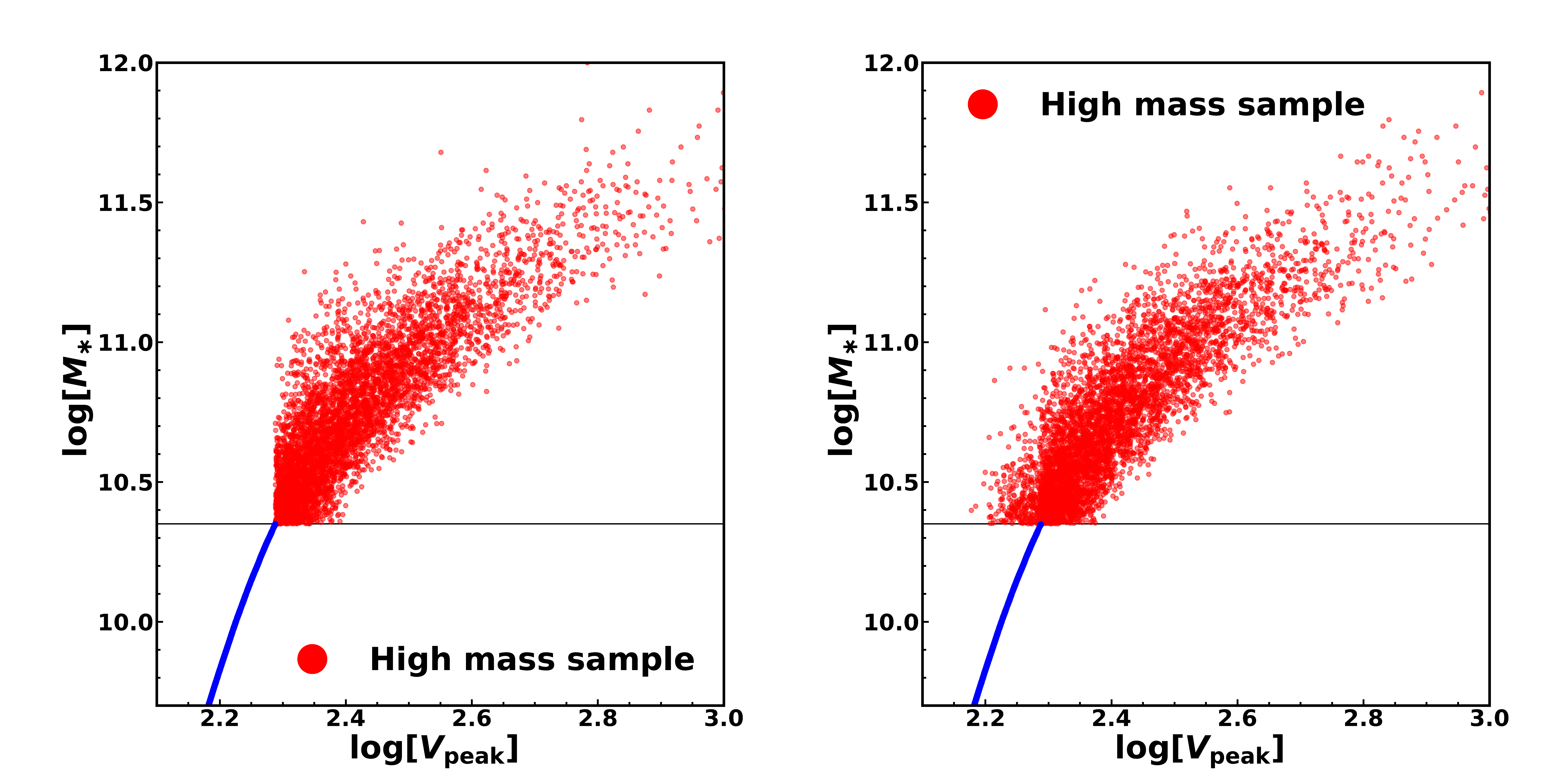}
\caption{Stellar mass as a function of $\vpeak$ for the two methods that only add scatter to high mass samples in \ref{wp_all_in_one_Vpeak_without_data_v1}. For both panel, we add $\sigma[\mgal|\vpeak]=0.2$. Red dots represent high mass samples and blue dots are other samples. There are two definitions of the high mass samples. The first one is to define high mass galaxies as galaxies bigger than $10^{10.35} \msol$($\log \vpeak=2.3$) after adding scatter. The second is to define high mass galaxies as galaxies with $\log \vpeak>2.3$. The first method makes sure $\sigma[\mgal|\vpeak]$ a constant at different $\vpeak$, but some low mass samples are scattered into high mass sample regime. The second method guarantee there won't be samples scattered to high mass regime from low mass regime, but will cause $\sigma[\mgal|\vpeak]$ smaller than the constant value around $\log \vpeak=2.3$. We note that the results of the test using both these samples give consistent results.}
\label{high_mass_only_v1}
\end{figure*}

Although the previous panel clearly shows that $\omega$ is sensitive to scatter, it is not clear what halo mass range the statistic is sensitive to. We performed a test to identify the halo mass scale by constructing mock samples with scatter in the central galaxies, but no scatter in the tracer galaxies. Specifically, we construct standard abundance matching models from which we select the central galaxy sample. At the $\vpeak$ values below the lower limit of the central galaxies, the tracer sample is selected by $\vpeak$ only, with limits adjusted to match the number of tracer galaxies. 
In the right-hand panel of Figure \ref{wp_all_in_one_Vpeak_without_data_v1}, we show the results of this test. Even though the value of $\slogmv$ for central galaxies increases, the value of $\omega$ is roughly constant. Thus, the trend in the left-hand panel is driven by the halos that contain galaxies with $\log\mgal=[9.7,10.35]$, or roughly $\mhalo\sim 10^{11.8}$ $\hmsol$.

To construct this test, a choice must be made about where to set the boundary between the part of the model with scatter and the part of the model with no scatter. The left-hand panel in Figure \ref{high_mass_only_v1}, this transition is set to be at a threshold in $\vpeak$. This method ensures a clear exclusion between the halos that contain central galaxies and halos that contain the tracers. However, the scatter in $\vpeak$ at fixed $\mgal$ deviates from the original lognormal, to some extent. The right-hand side shows a model in which the transition is set to be a constant value of $\mgal$. In this model, the distribution of $\vpeak$ is preserved, but there is some overlap in the masses of the halos used in the central sample and the tracer sample. However, both of these complementary approaches yield the same results, with $\omega$ no longer being dependent on scatter.

An important caveat for the $\omega$ statistic is that scatter is not the only thing that can impact $\fsat$ in abundance matching. While we limit ourselves to abundance matching models of single halo properties, \cite{lehmann2016concentration} constructed an abundance-matching proxy by combining halo mass and halo concentration into a single quantity, with a free parameter that allowed variations of the relative importance of each parameter in the rank-ordering of the halos. In their results, the satellite fraction depended on the relative importance of concentration at fixed scatter, while also being able to fit observed clustering measurements. Thus our constraints on $\slogmv$ from $\omega$ do not take this into account. However, our second statistic, which we describe presently, does not depend on $\fsat$ and thus does not have this concern. 

\subsection{Line-of-sight velocity distribution around central galaxies}\label{sec:Line-of-sight}

We define the velocity of a satellite galaxy ($v_{\rm sat}$) as the relative velocity between the satellite and then central galaxy within the same halo. Sub-halo velocities in $N$-body simulations have been shown to follow a Gaussian distribution at fixed subhalo $M_{\rm peak}$
(\citealt{faltenbacher2006velocity}). 
If there is no scatter between $\mhalo$ and $\mgal$, the satellite galaxies around centrals at fixed stellar mass would also have a Gaussian distribution. As scatter increases, the distribution of halo mass at fixed stellar mass widens. Thus, the distribution of $v_{\rm sat}$ will be a sum of multiple Gaussians of varying widths, yielding a distribution function with positive kurtosis, $\kappa$. The larger the scatter is, the higher $\kappa$ will be. The variance of the velocity distribution will also change with scatter, but the variance is sensitive to the overall mass scale of the SHMR, as well as the cosmology chosen. Thus, $\kappa$ is a more reliable diagnostic for scatter. 

\begin{figure*}
\includegraphics[width=\textwidth]{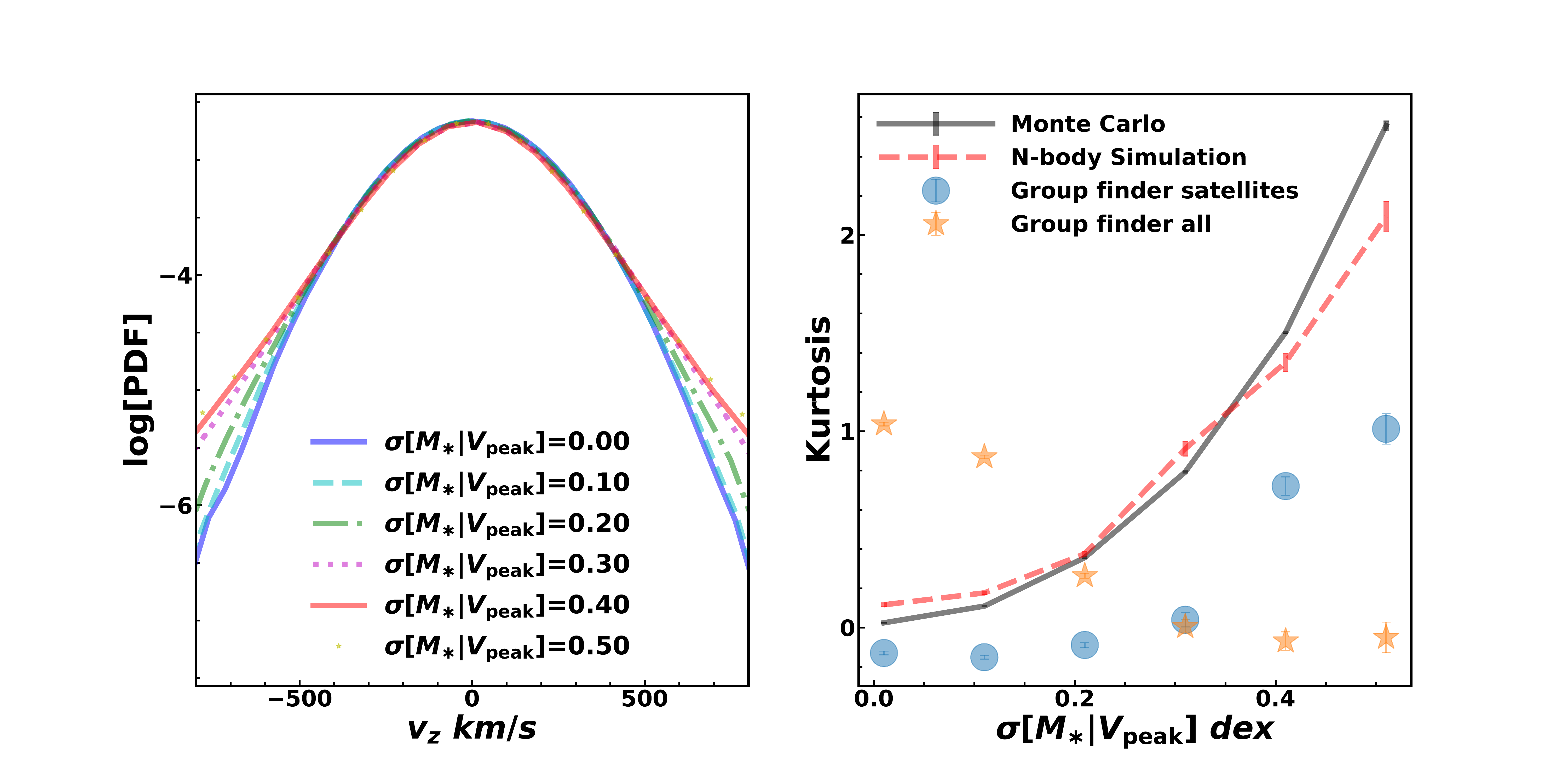}
\caption{The left panel shows the probability density function of velocities from virial theory under different scatters, calculated by Monte Carlo. As the scatter gets larger, the distributions broaden due to the wider host halo mass range. The right panel compares results of the kurotsis from Monte Carlo and abundance matching mocks. The black line is from the Monte Carlo. A set of host halo around $10^{12} \msol$ are generated, and velocities of satellites for these host halos are produced by virial theorem. To compare properly to the simulation results, we convert the scatter $\sigma[\mgal|\vpeak]$ to scatter in halo mass at $\log[\mgal]$ around 10.5 (or $\log \mgal$ around 12.2. The red line shows the results from our abundance matching models, centered at the same stellar mass. The symbols are mock results after passing them through the group findre. The blue dots uses group-finder satellites to measure the $v_z$ PDF. The yellow stars us all tracer galaxies. All error bars are derived using bootstrap resampling of the sample of central galaxies.}
\label{vz_from_MCMC_v4}

\end{figure*}

In Figure \ref{vz_from_MCMC_v4}, we check the assumptions made in the previous paragraph. Namely, a Gaussian distribution of both the satellite velocities and the distribution of halo mass at fixed stellar mass. The left-hand panel shows the probability distribution function (PDF) of line-of-sight satellite velocities, $v_z$, within host halos at fixed central $\mgal$. The velocity distribution within each host halo is assumed to be a Gaussian with a position-independent dispersion set by the virial theorem. The left-hand panel shows the PDF of $v_z$ as scatter increases. In this calculation, we assume the distribution of $\mhalo$ is a log-normal, but in the key we show the value of $\slogmv$ that corresponds to the scatter in the calculation, using the conversion tables described in the previous section. Although this calculation is analytically tractable, we use the Monte Carlo method
to construct the PDF by randomly sampling from both the distribution of host halo masses and distribution of subhalo velocities. This allows us to test the impact of finite sample size on the estimation of $\kappa$. Although using the full PDF creates the strongest dependence of $\kappa$ on $\slogmv$, the small number of velocities in the wings of the distribution create noise in the estimation of $\kappa$. In the balance between correlation strength and precision with which $\kappa$ can be measured, we find that estimating $\kappa$ after removing the regions of the PDF outside of $\pm 2\sigma_v$ yields the best results. In the right-hand panel, we show how $\kappa$ (with $2\sigma_v$ clipping) increases with $\slogmv$. Even with many simplifying assumptions in this calculation, the assumption of Gaussian subhalo velocities and lognormal halo mass distribution agrees very well with the $\kappa$ results from our abundance matching mocks.

The analytic and abundance matching results described above assume perfect knowledge of what is a central and what is a satellite galaxy. In survey data, there is a background contribution of interlopers that are not actually associated with the central galaxy, but are in fact centrals in nearby halos projected along the line of sight. The blue dots in the right hand panel of Figure \ref{vz_from_MCMC_v4} show how $\kappa$ depends on scatter when using the PDF of {\it all} galaxies in a cylinder of width $\Delta v_z = 1000\,\kms$ around each central galaxies (but still employing $2\sigma_v$ clipping). Surprisingly, $\kappa$ actually decreases with the increasing of scatter. This is due to the shape of the background $v_z$ distribution, which is broad and roughly consistent with a Gaussian. When scatter is zero, the $v_z$ PDF of true satellites is also a Gaussian, and the resulting combination of the two distributions yields a positive $\kappa$. As the scatter increases, the wings of the $v_z$ PDF for true satellites become significant, and the resulting combination with the broad background has {\it less} kurtosis. However, this result shows that $\kappa$ for all galaxies, without any background subtraction, has constraining power on scatter, especially when the scatter is low. 

Removing the background is a not a trivial task, however.  The yellow stars in the right-hand panel show how $\kappa$ depends on scatter when using the group finder results to estimate the background. The results show $\kappa$ only for galaxies identified as satellites by the group finder. We once again recover the monotonic relation between $\kappa$ and $\slogmv$, but now the correlation is much weaker than the results using true satellites. This is because the group finder places a Gaussian prior on the velocity distribution of satellites at fixed halo mass. Thus, if the group finder's estimate of halo mass is biased, this will also bias the PDF of $v_z$. At low group masses, where the number of satellites per halo is low, the group finder has a tendency to underestimate the scatter in halo mass at fixed stellar mass (\citealt{reddick2013connection}). Thus, this will reduce the $\kappa$ at fixed stellar mass, as is shown in Figure \ref{vz_from_MCMC_v4}. 

The results in Figure \ref{vz_from_MCMC_v4} do show, however, that the $v_z$ PDF of all galaxies and the PDF of group catalog satellites offer complementary constraints on scatter: the $\kappa$ for all galaxies is sensitive to the scatter at low scatter value, while the $\kappa$ for group finder satellite is sensitive at high scatter value. Thus, we use both quantities to constrain scatter.

\begin{figure*}
\includegraphics[width=\textwidth]{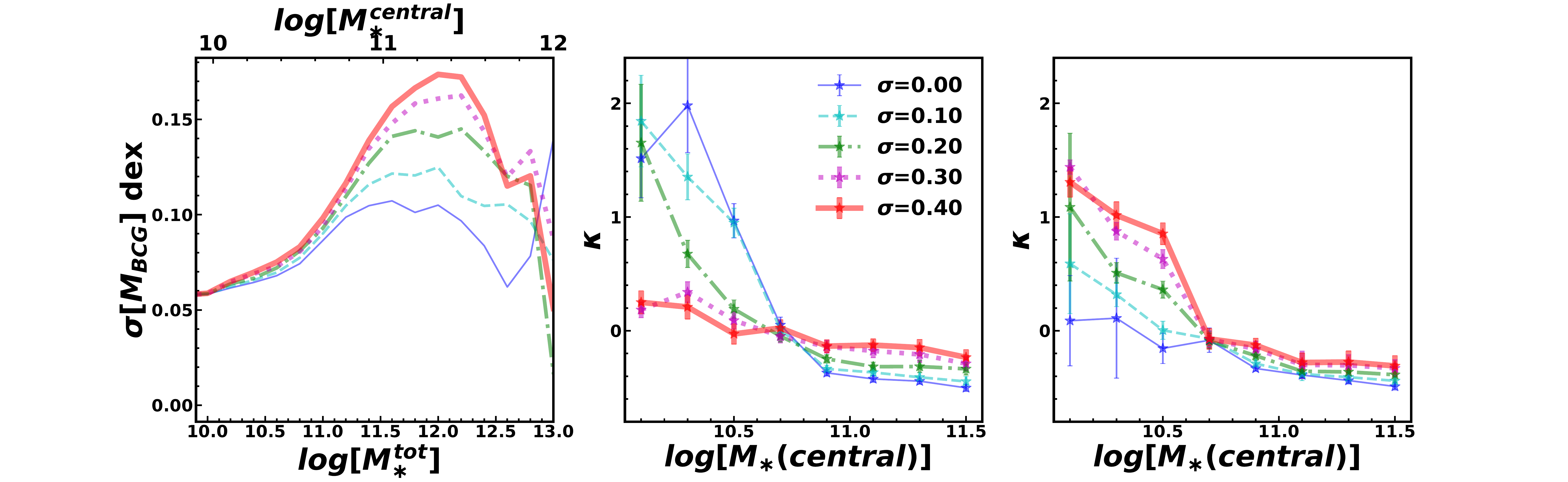}
\caption{Left-hand panel: The scatter in $\mgal$ of central galaxies identified by the group finder, plotted as a function of the total stellar mass in the group. This is a reproduction of the method used by \protect\cite{reddick2013connection} to constrain $\slogmv$, showing that the models differentiate at halo mass scales above $\sim 10^{13}$ $\msol$. Middle panel: The kurtosis method, applied to all tracer galaxies, as a function of central $\mgal$. Right panel: The kurtosis method, applied to group-finder satellites, as a function of central $\mgal$. In both of these panels, the models converge at high masses, while differentiating at low masses. This shows the complementarity of our approach to that of \protect\cite{reddick2013connection}, even though the data are the same.}
\label{Kurtosis_at_differetn_central_mass_fusion_v3}
\end{figure*}

\cite{reddick2013connection} used the same group catalog to constrain scatter. Their method was to use the scatter in $\log \mgal$ of central galaxy as a function of total group stellar mass, which the group finder uses as its proxy for halo mass. Figure \ref{Kurtosis_at_differetn_central_mass_fusion_v3} shows that our use of the $\kappa$ in $v_z$ is complementary to their method. The left panel shows our reproduction of the Reddick theoretical results using our abundance matching models, which are based on a different stellar mass function and modified abundance matching method. The constraints on scatter from this method are derived at halo mass scales where the central galaxies are bigger than $10^{11} \msol$, corresponding to halo mass bigger than $10^{13} \msol$. The middle and right panels show $\kappa$ as a function of central galaxy stellar mass for different values of $\slogmv$. These quantities---both using all galaxies and using only group finder satellites---are more sensitive to scatter at $\mgal \lesssim 10^{10.5} \msol$, $\mhalo \lesssim 10^{12} \msol$.

\begin{figure*}
\includegraphics[width= \textwidth]{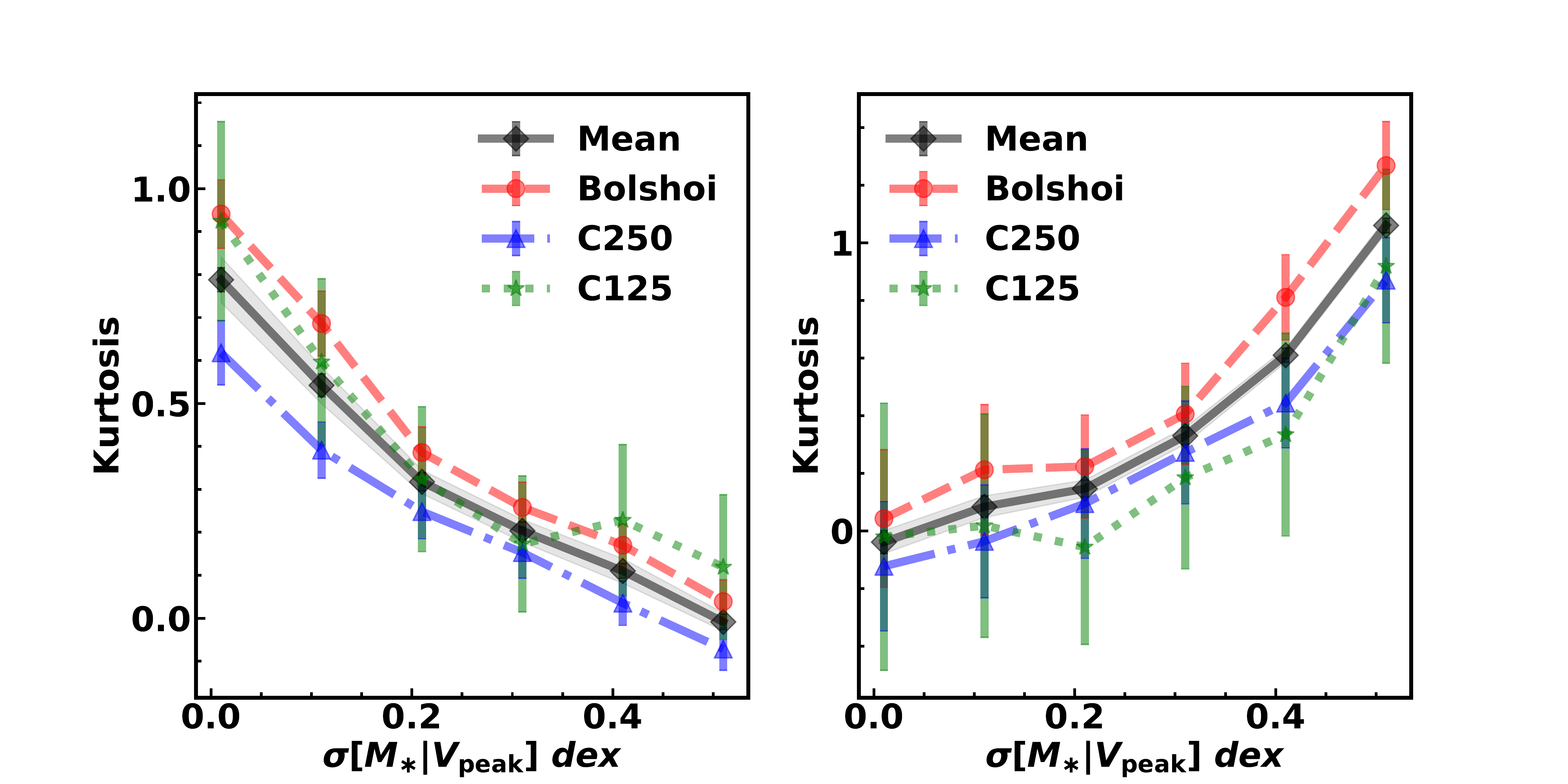}
\caption{Kurtosis ($\kappa$) as a function of scatter in the abundance matching models applied to simulations. The black line is the average from the three simulations and errors are from bootstrap resampling. A two-sigma clipping is applied to avoid the noisy tails at high velocities. The left panel shows $\kappa$ as a function of scatter for all tracer galaxies. The right panel shows kurtosis from tracer galaxies identified as satellites by the group finder.}
\label{Kurtosis_signal_all_whole_box_without_data_v1}
\end{figure*}

Figure \ref{Kurtosis_signal_all_whole_box_without_data_v1} shows our theoretical results, comparing $\kappa$ for all three simulations. Here, we use the same central galaxies as defined for the $\omega$ statistic. The black symbols show the volume-weighted mean of all simulations. Once again, the results from the higher-resolution C125 are consistent with the lower resolution simulations.

\section{Results}\label{sec:results}

We compare the predictions for our mock catalogs to the measurements from the NYU VAGC catalog. Recall that all comparisons between theory and observations are performed after the group finder has been applied to the mock galaxy catalogs, with the sample sample selections used for central and tracer samples.

\subsection{\texorpdfstring{$w_p$}{wp} ratio}\label{sec:ratio_result}

\begin{figure}
\includegraphics[width=\columnwidth]{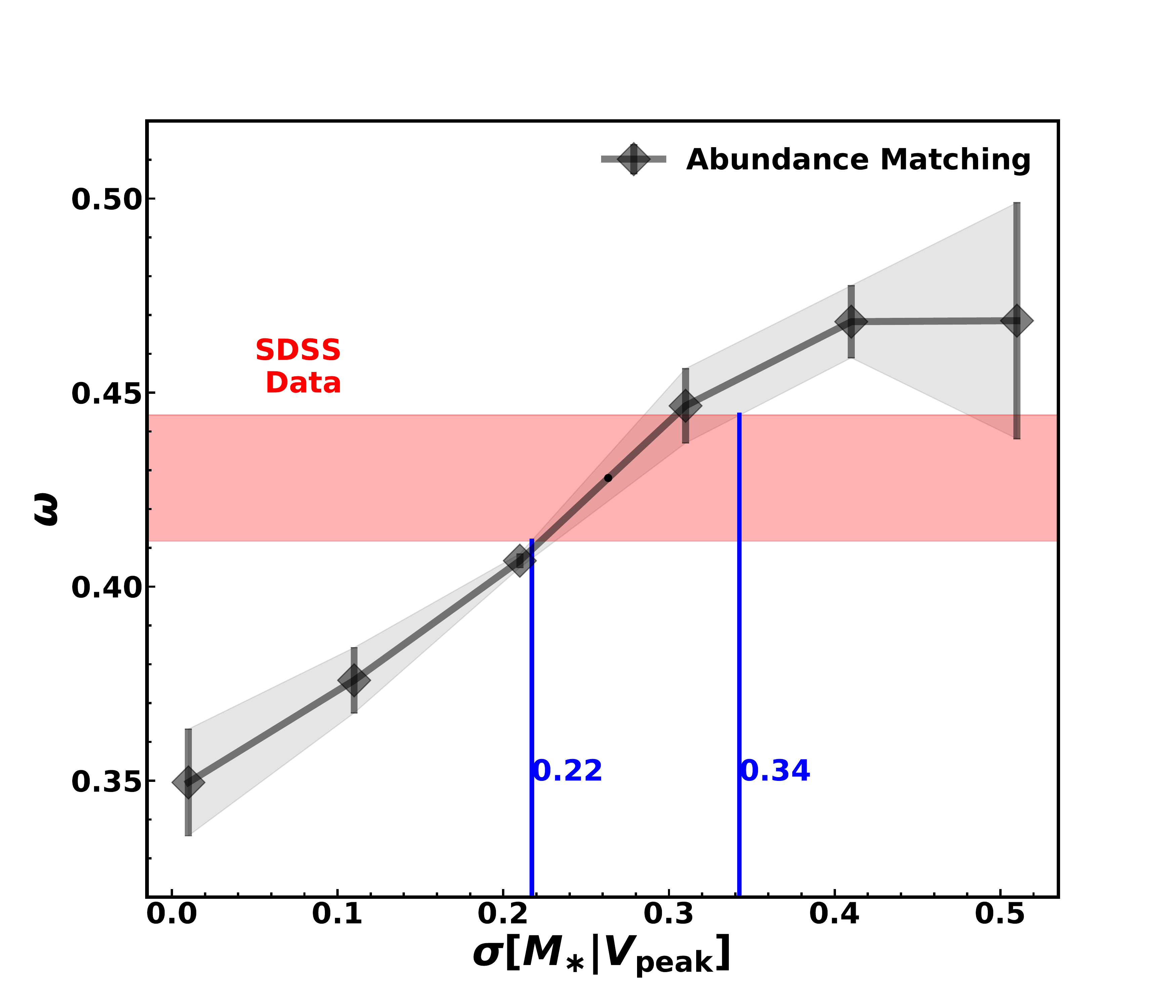}
\caption{Ratio of auto to cross correlation $\omega$, 
in the SDSS group catalog compared to theoretical models. 
The ratio calculated from the SDSS group catalog is the red shaded area, and the error bar is estimated using the jackknife method. The black line indicates the average of the results of mock catalogs constructed from three simulations; the error (shaded region) is taken to be the difference between the C250 and BolshoiP simulations. The blue lines represent the range of scatter values from the intersection of the measurements with the predictions. The scatter $\sigma[\mgal|\vpeak]$ for tracer galaxies is between 0.22 and 0.34 ($\log M_{\ast}$ between 9.7 and 10.35). }
\label{wp_all_in_one_Vpeak_fusion_v1}
\end{figure}

Figure \ref{wp_all_in_one_Vpeak_fusion_v1} compares our measurement of $\omega$ from the SDSS group catalog to our simulations. Using the samples defined in Table 1, we find $\omega=0.428\pm 0.016$, where the observational error-bar was determined through jackknife resampling on the plane of the sky using 25 total subsamples. The solid curve shows the mean of our models from the three $N$-body simulations, with the error indicated by the shaded area.  Using the lower and upper one-sigma error range in the mock mean to convert this measurement to a one-sigma range in scatter. This yields a constraint of $\sigma[\mgal|\vpeak] =  0.275\pm 0.055$. 

\subsection{\label{sec:Line-of-sight_result}Line-of-sight velocity distribution around central galaxies }

\begin{figure}
\includegraphics[width=\columnwidth]{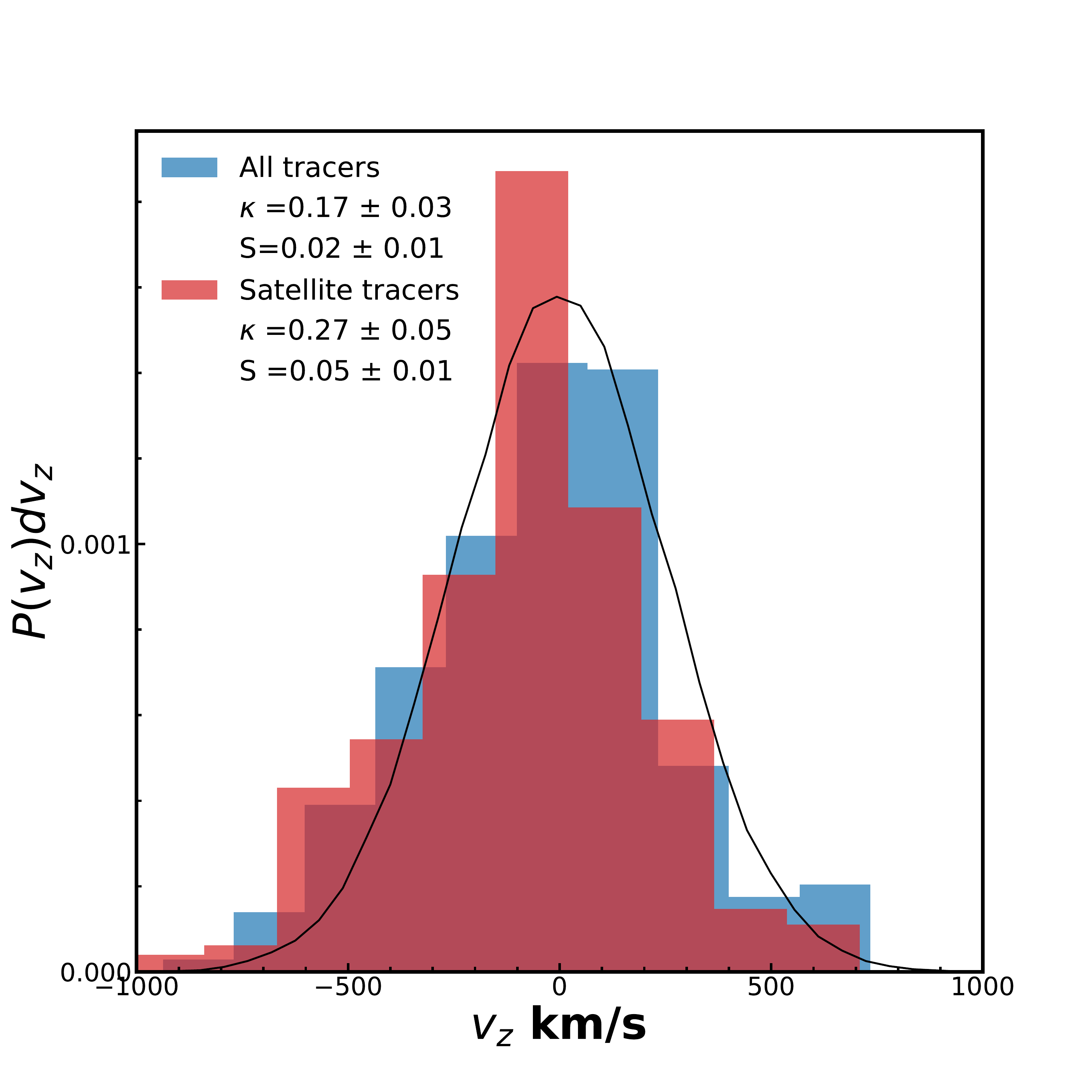}
\caption{Histogram of velocities around central galaxies in the SDSS group catalog. To compute $\kappa$, two-sigma clipping is applied, as is done with the mocks. The black line is a Gaussian distribution (thus zero kurtosis). The blue histogram shows the velocity distribution for satellite velocities of all galaxies inside the cylinder. The kurtosis, $\kappa$, is equal to 0.17 and skewness, $S$, is consistent with zero. The red histogram is the velocity distribution for galaxies labeled as satellites by the group finder. For this sample, $\kappa=0.27$, and there is some skewness in the sample. There are 31,923 galaxies; 8,966 of these are identified as satellites.}
\label{PDF_vz_VAGC_NYU_paper_v2}
\end{figure}

Figure \ref{PDF_vz_VAGC_NYU_paper_v2} shows the distribution of of $v_z$ around central galaxies in the SDSS group catalog. The blue line is the PDF for all tracer galaxies, while the red line shows the PDF for group-identified satellites. The relative areas under the two curves represents the different total number of galaxies in the `all' and `group satellites' subsamples of the overall tracer population. The key in the figure also indicates the kurtosis values found in each PDF, with errors determined by bootstrap resampling of the sample of central galaxies. As discussed in the previous section, all these results include $2\sigma_v$ clipping of the PDF to calculate $\kappa$.

\begin{figure*}
\includegraphics[width=\textwidth]{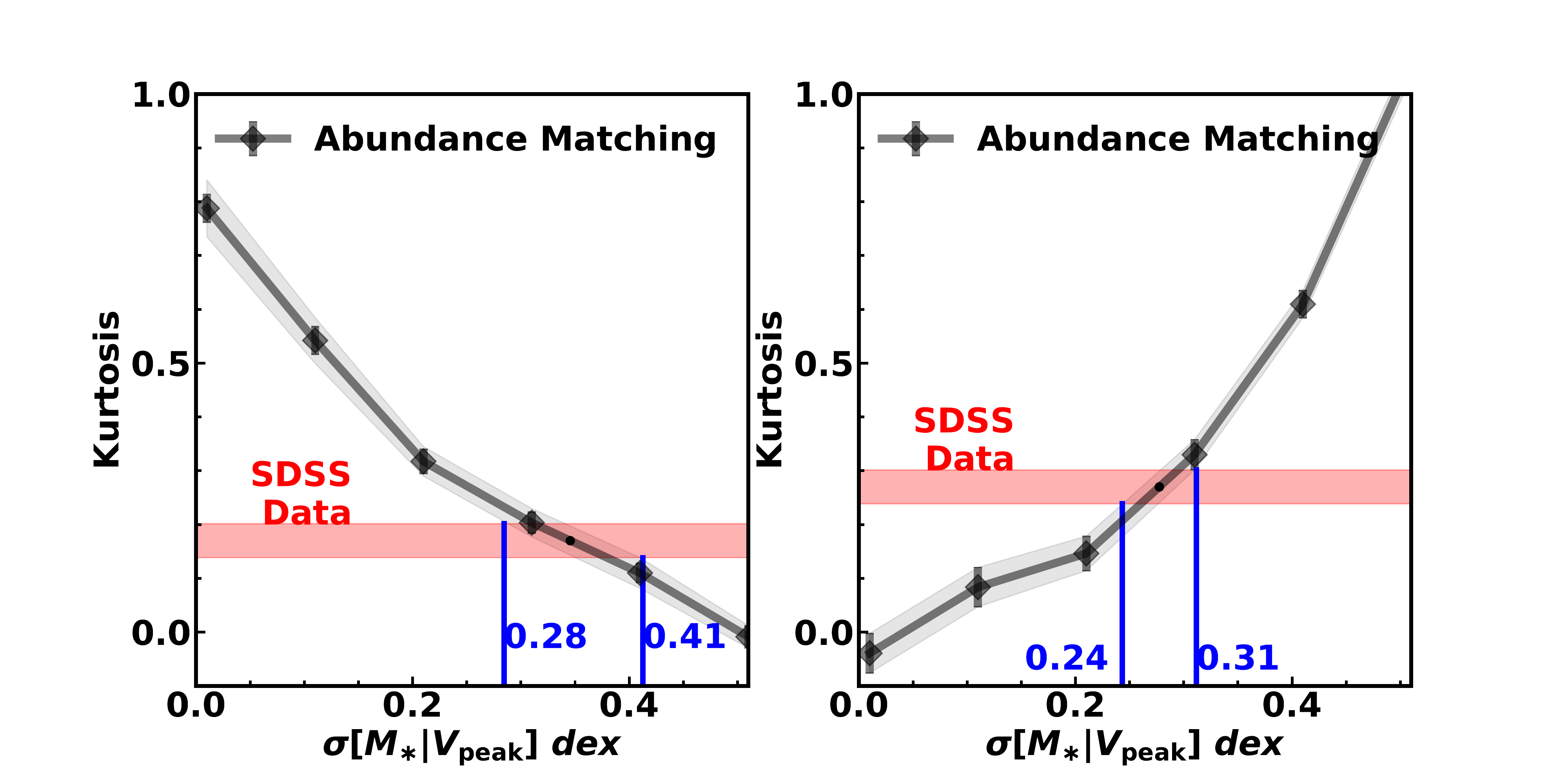}
\caption{Kurtosis of our models compared to the kurtois measured from the SDSS group catalog. The black line is the average from all three simulations. The red shaded bands are the $\kappa$ values from SDSS. In the left panel we show the results using all tracer galaxies. The SDSS value is $\kappa=0.17\pm 0.02$, yielding a scatter between 0.28 and 0.41. In the right panel, we show the results using the group satellites in the tracer sample. The SDSS value is $\kappa=0.27\pm 0.02$, yielding a scatter between 0.24 to 0.31.  Combining these two results yields $\sigma [\mgal|\vpeak] = 0.30 \pm 0.03$ ($\log M_{\ast}$ between 10.35 and 10.65).}
\label{Kurtosis_signal_all_whole_box_v1}
\end{figure*}

Figure \ref{Kurtosis_signal_all_whole_box_v1} compares the SDSS $\kappa$ measurements to our model predictions from Figure \ref{PDF_vz_VAGC_NYU_paper_v2}. We use the upper and lower $1\sigma$ error bars on the mean theoretical prediction to set the $1\sigma$ ranges on the $\slogmv$ constraints. The scatter constrained from all tracer galaxies is $\sigma [\mgal|\vpeak] = 0.35 \pm 0.06$, while $\sigma [\mgal|\vpeak] = 0.27 \pm 0.04$ for satellite tracer galaxies, a difference of $\sim 1$ sigma. Combining the two results yields a final result of $\sigma [\mgal|\vpeak] = 0.30 \pm 0.03$. 

\subsection{Combined results}

\begin{figure*}
\includegraphics[width= \textwidth]{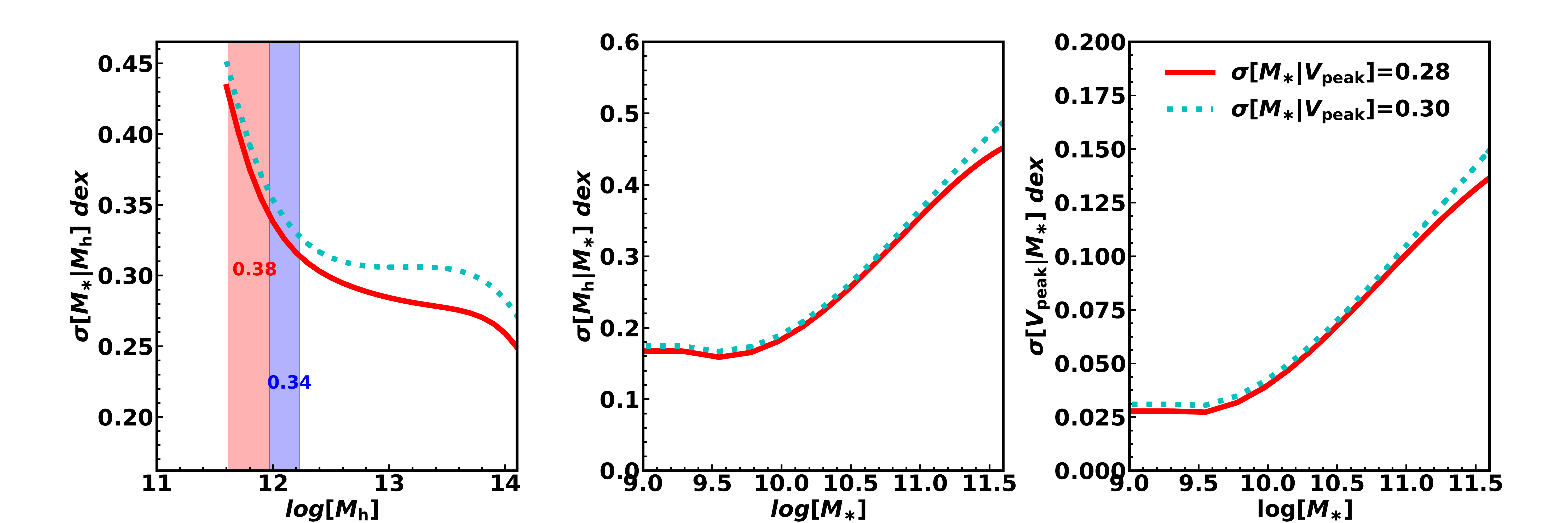}
\caption{Our SDSS constraints, translated into different definitions of scatter. The red and the cyan lines are made by using the scatter results derived from our $\omega$ and $\kappa$ techniques. The $\omega$ statistic constrains scatter at stellar mass between 9.7 and 10.35 (tracer galaxies). Converting the scatter from $\vpeak$ to $\mhalo$ yields $\sigma[\mgal|M_{\rm h}]$=0.38. The cyan line is from the $\kappa$ technique, isolating stellar masses between 10.35 and 10.65 (central galaxies). The scatter in $\sigma[\mgal|M_{\rm h}]$ is 0.34.}
\label{scatter_with_data_v1}
\end{figure*}

Figure \ref{scatter_with_data_v1} presents our constraints of $\slogmv$ translated into other quantifications of the scatter. In each panel, the two lines show the constraints from our two different methods of measuring scatter. The left-hand panel shows our models, which are constant $\slogmv$ at all values of $\vpeak$, in terms of $\sigma[\mgal|\mhalo]$. The halo mass ranges probed in each method are also shown with the vertical shaded regions. The two method yield results are consistent within their error bars. The $\sigma[\mgal|\mhalo]$ values are roughly constant at masses $\mhalo\gtrsim 10^{12.3}$ $\msol$. At lower halo masses, the scatter rapidly increases due to the scatter between $\vpeak$ and $\mhalo$, as shown in \S 3. The other two panels show the scatter when binning by $\mgal$ rather than halo properties. In this projection, the scatter is constant at $\mgal\lesssim 10^{10} \, \msol$, and monotonically increases at higher stellar masses. 

\section{\label{sec:Discussion}Discussion}

\begin{figure*}
\includegraphics[width= 130mm]{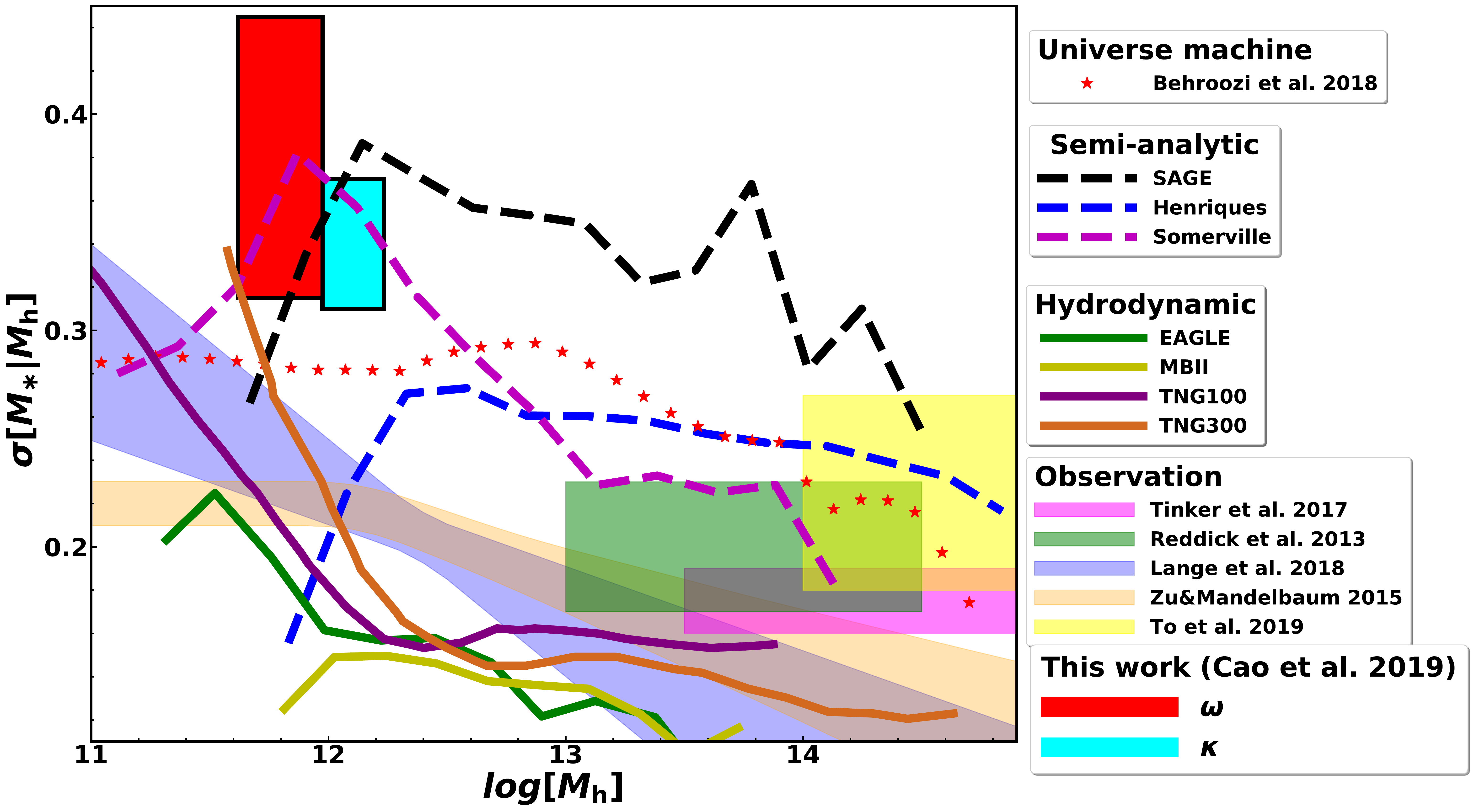}
\caption{We compare the scatter predicted in hydrodynamic and semi-analytic models of galaxy formation to constraints obtained observationally. Our results are the two shaded regions with black outlines, representing the only independent constraints at $\sim 10^{12}$ $\msol$. The two observational methods that parameterize scatter as being mass-dependent both yield scatters that are larger at lower halo masses. Although the overall amplitude of the scatter is higher in the data than in the observational results, the consensus of the data agree qualitatively with the mass-dependence of scatter predicted by hydrodynamical models.}
\label{scatter_compare_v1}
\end{figure*}

We have presented two new methods to constrain the scatter in the relationship between galaxy stellar mass and their host dark matter halos. These methods probe galaxy and halo mass scales that have not been independently probed before. Our model to connect galaxies to halos uses the peak historical value of the halo's maximum circular velocity, $\vpeak$, as the halo property that is matched to stellar mass. We use this property because it more accurately reproduces galaxy clustering measurements, in comparison to halo properties based on halo mass (\citealt{reddick2013connection}). 

As discussed in \S 4.1, abundance matching modeling can be constructed from more than one halo property. \cite{lehmann2016concentration} use the quantity $V_\text{vir,peak}\left(V_\text{max,peak}/V_\text{vir,peak}\right)^\alpha$ as their abundance matching proxy, with the free parameter $\alpha$ determining the relative importance of halo mass and halo concentration in halo rank-ordering. Using this proxy, $\alpha=1$ is equivalent to the $\vpeak$ model used here. This model allows the galaxy satellite fraction to vary at fixed values of scatter, which our $\omega$ results do not take into account. For $L\sim L_\ast$ galaxies, \cite{lehmann2016concentration} found $\alpha \approx 0.5$ produced the best fit to observed clustering. Decreasing $\alpha$ decreases $\fsat$, which increases the value of $\omega$ and would yield a lower constraint on scatter given the observed value of $\omega$. The kurtosis method, which does not depend on $\fsat$, would not change. A discrepancy between the values of $\slogmv$ obtained with the $\omega$ and kurtosis methods might imply $\alpha$ deviating from unity. With our current statistical precision, the results from the $\omega$ statistic are in agreement with the results from the kurtosis method, although the constraints on scatter from $\omega$ are significantly weaker than from kurtosis. But more than demonstrating agreement between the two methods, if future data can reduce the uncertainties on $\omega$, such a comparison would offer a method to further constrain $\alpha$ in the \cite{lehmann2016concentration} model. 

Figure \ref{scatter_compare_v1}, a variation of Figure 8 in \cite{wechsler2018connection}, compares our measurements to existing measurements, as well as to predictions from different theoretical models of galaxy formation. The shaded regions indicate observational constraints on scatter as a function of $M_{\rm h}$. The clustering-inferred results from \cite{tinker2017correlation} only use galaxies with $\mgal\gtrsim 10^{11.4}$ $\msol$, thus are relegated to halo masses above $\sim 10^{13}$ $\msol$. Figure \ref{Kurtosis_at_differetn_central_mass_fusion_v3} demonstrated that the group catalog results of \cite{reddick2013connection} come from $\mhalo\gtrsim 10^{13}$ $\msol$. We also include the analysis of RedMapper clusters of To et.~ al.~(in preparation). The satellite kinematics results of \cite{lange2018updated} parameterize scatter as a power-law function of halo mass, but the constraints are weighted to halos with the highest number of satellites, and also include galaxy clustering in the analysis as well. \cite{zu2015mapping} use both galaxy clustering and weak lensing to constrain a different parametric function for $\sigma [\mgal]$. Their model is a power-law dependence on $\mhalo$ at high halo masses, with the scatter at low masses fixed to a constant with the value of the power law at the break point. All of these constraints are generally consistent, with values of $\sigma [\mgal|\mhalo] \lesssim 0.2$ dex\footnote{We note that the constraints of \citealt{lange2018updated} are of the scatter in luminosity, rather than stellar mass. Physically, we expect the scatter in luminosity to be larger than stellar mass due to the scatter of $L$ at fixed $\mgal$. But observational errors are on $\mgal$ are significantly higher, thus these two scatters are generally found to be roughly consistent with one another.}. 

The values that we constrain for $\sigma [\mgal|\mhalo]$---which are converted from our constraints on $\sigma [\mgal|\vpeak]$---are significantly higher, with $\sigma [\mgal|\mhalo] \sim 0.35$ dex. Our constraints are at significantly lower halo masses as well, spanning the range $\mhalo=[10^{11.6},10^{12.0}]$ $\msol$ 
The galaxies probed in our study are significantly different than those that occupy $10^{13}$ $\msol$ halos, which are mostly red and passive.  In contrast, the quiescent fraction of central galaxies in the halos we probe is 0.3. The observational results that overlap the halo mass range probed here are extrapolations from higher mass to lower mass, where most of the constraining power lies. For both analyses, a strong upturn at $\mhalo\sim 10^{12}$ $\msol$ would not be consistent with their functional forms.

However, the scenario implied by the combination of previous results with our new results---in which scatter is quite small for massive halos, but rapidly increases below the pivot point in the stellar-to-halo mass relation---is consistent with the results of hydrodynamic simulations. Figure \ref{scatter_compare_v1} shows results for the EAGLE, IllustrisTNG, and MassiveBlackII (MBII) simulations (\citealt{crain2015eagle}, \citealt{springel2017first}, \citealt{khandai2015massiveblack}). Both TNG100 and TNG300 show a significant upturn in $\sigma [\mgal|\mhalo]$ at low halo masses. EAGLE shows a less pronounced upturn at a similar mass scale. MBII shows no upturn, but resolution limits on the MBII simulation prevent exploration of mass scales significantly below the pivot point. 

Figure \ref{scatter_compare_v1} also shows several semi-analytic models, as well as results of the forward empirical model Universe Machine (\citealt{behroozi2018universemachine}), all of which are significantly higher than the hydrodynamical models, but have less dependence on halo mass. \cite{mitchell18} compared hydrodynamic and semi-analytic models applied to the same dark matter simulations, finding significant scatter in the stellar masses between the two models. This scatter, they concluded, was created in the semi-analytic models due to differences in the treatment of gas accretion and its correlation with AGN feedback. The predictions of Universe Machine, which are driven mainly by the different mass accretion histories of halos at fixed $\mhalo$, are consistent with the semi-analytic models.

In the data, as well as the hydrodynamic models, the mechanism that drives the decrease in scatter at high masses can be explained in part by the halo property chosen to express the scatter. If galaxy stellar mass is more correlated with $\vpeak$ than $\mhalo$, then the $\sigma [\mgal|\mhalo]$ will naturally increase at $\mhalo\lesssim 10^{12}$ $\msol$. However, even quantified as $\sigma [\mgal|\vpeak]$, our results show a marked increase in scatter at low $\mhalo$, starting at $\sigma [\mgal|\vpeak] \sim 0.18$ at high masses, and $\sim 0.27-0.30$ for our results here. 

But beyond abundance-matching bookkeeping, the physics of star-formation and quenching likely plays a role in determining any mass dependence of the scatter. Empirical models constrained to match the width of the star-forming main sequence predict $\sigma [\mgal|\mhalo] >0.2$ dex, regardless of how star formation histories vary about the mean star formation rate, or the degree of correlation between halo mass growth and stellar mass growth (Hahn et al, in prep; \citealt{behroozi2018universemachine}). The value of the scatter at high masses, however, is sensitive to the mechanism that triggers galaxy quenching (\citealt{tinker2017testing}). If quenching is triggered at a galaxy mass threshold, then the scatter of $\mgal$ at fixed $\mhalo$ is attenuated, regardless of the amplitude of the scatter before the quenching occurs. 

\section{Summary}

In this paper, we have developed two new methods for measuring the scatter between halos and galaxies that are sensitive to specific mass ranges in the galaxy--halo connection: (1) the ratio of the auto correlation of central galaxies to the cross correlation of these galaxies with tracer galaxies (``cross-correlation technique''), and (2) the kurtosis of the pairwise velocity distribution around central galaxies (``kurtosis technique'').  These methods are most sensitive to 
galaxies that occupy in $10^{12}$ $\msol$ halos, the halo mass in which most star formation occurs; this mass scale is lower than that probed by two-point galaxy clustering and group statistics, which have been the most commonly used methods to constrain scatter in the galaxy--halo connection.  We have applied these methods to a samples of central galaxies constructed from the SDSS main galaxy sample. Our principle results are as follows:

\begin{itemize}
    \item We find $\slogmv=0.27\pm 0.05$ dex from the cross-correlation technique and $\slogmv=0.30\pm 0.03$ dex from the kurtosis technique. 
     \item These results are consistent with each other but are significantly higher than values obtained at higher halo masses, which generally lie at $\slogmv<0.2$ dex.
     \item Due to scatter between halo $\vpeak$ and $\mhalo$, our measured values of scatter are even higher when converting to $\slogmm$.
     \item The rapid increase in scatter at $\mhalo\approx 10^{12}$ $\msol$ is qualitatively consistent with predictions from hydrodynamic simulations, but with the observations showing larger scatter values at all $\mhalo$.
\end{itemize}

The results presented here can be extended with near-term survey data. The Bright Galaxy Survey of the Dark Energy Spectroscopic Instrument (\citealt{desi_fdr}) will expand the SDSS main galaxy sample two magnitudes fainter and cover twice the area. Use of this data will both tighten the constraints at Milky Way masses as well as provide constraints a significantly smaller halo and stellar masses. Current high-redshift surveys, such as DEEP2 (\citealt{deep2}) and zCOSMOS (\citealt{zcosmos}), provide SDSS-like selections of galaxies at $z\sim 1$, allowing for the detection of any redshift evolution in scatter at our target mass range. Combined, these data and the constraints on scatter they would yield represent and excellent test of galaxy formation models, in addition to providing critical information for our knowledge of the galaxy-halo connection.


\section*{Acknowledgements}

The simulations used in this study were produced with computational resources at SLAC National Accelerator Laboratory, a U.S.\ Department of Energy Office; YYM and RHW thank Matthew R. Becker for creating these simulations, and thank the support of the SLAC computational team. 
YYM was supported by the Samuel P.\ Langley PITT PACC Postdoctoral Fellowship, and by NASA through the NASA Hubble Fellowship grant no.\ HST-HF2-51441.001 awarded by the Space Telescope Science Institute, which is operated by the Association of Universities for Research in Astronomy, Incorporated, under NASA contract NAS5-26555.

\bibliography{citation} 

\appendix
\section{Implementation of Abundance Matching}

\begin{figure*}
\includegraphics[width=\textwidth]{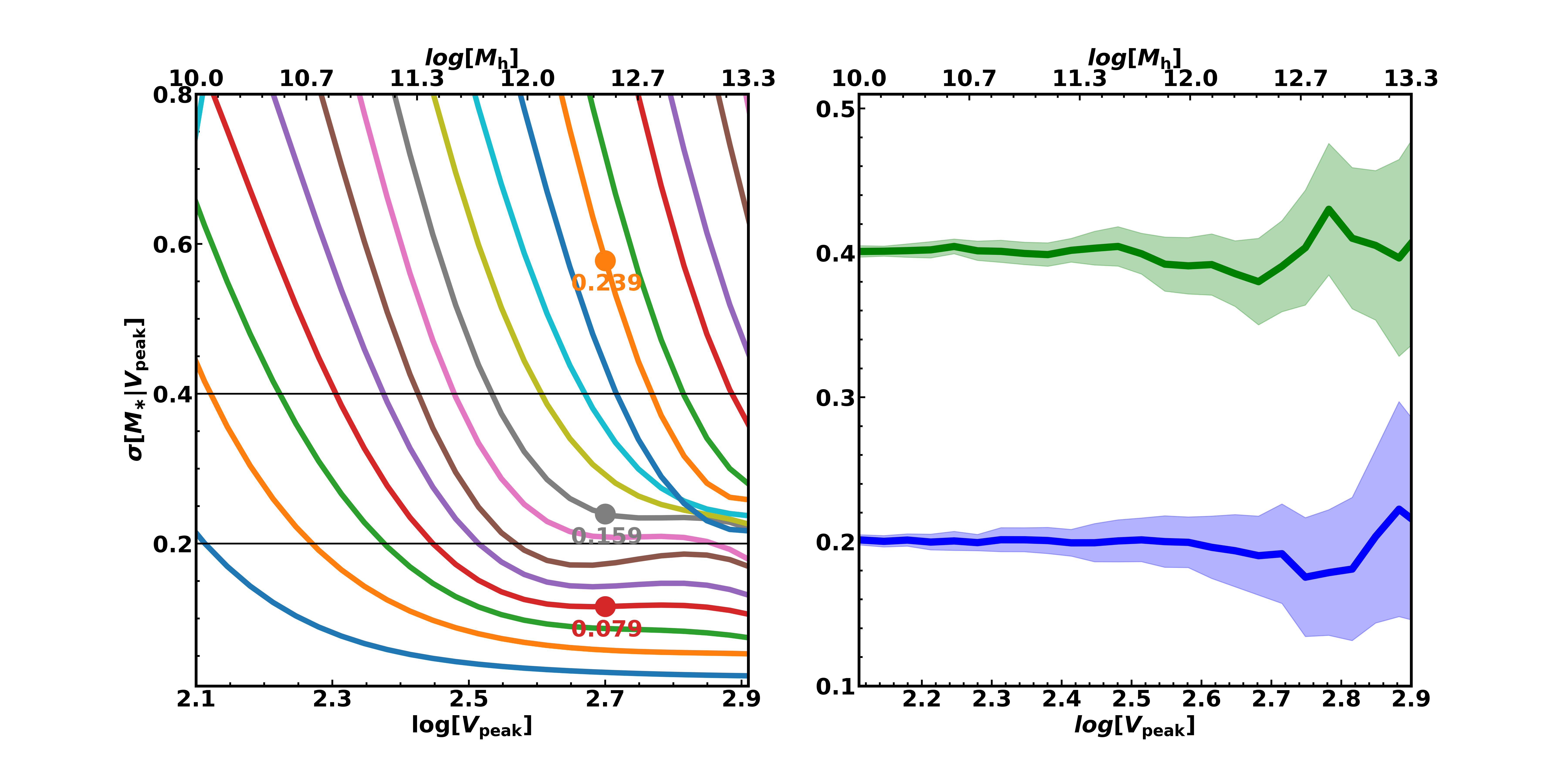}
\caption{Left panel: $\sigma[ \mgal|\vpeak]$ as a function of $\vpeak$. This figure shows the results of a suite of models with different values of $\sigma(\log  \vpeak)$. Each curve represents a model with a different $\sigma(\log  \vpeak)$, which is a constant for all halos. For context, we show the values of $\sigma(\log\vpeak)$ for three models, which are indicated by large dots at $\log\vpeak=2.7$. The two horizontal lines show $\sigma[ \mgal|\vpeak] =0.2$ and $0.4$ are added as reference to show what $\slogvm$ is required to be at each $\vpeak$ to make $\slogmv$ a constant. Right panel: The blue and green line represent $\sigma [\mgal|M_{\rm h}]$ as a function of $\vpeak$ by using our method of adding $\sigma [\mgal|M_{\rm h}]$=0.2 and 0.4. Errors are estimated as the variance in 100 independent realizations at each scatter. }
\label{scatter_tabulated_v1}
\end{figure*}

Before creating individual mocks, we first construct a table of models, as presented in Figure \ref{scatter_tabulated_v1}. This figure shows the results of a suite of models with different values of $\sigma(\log\vpeak)$, varied from 0 to 0.5 in 500 linearly spaced bins. Because this scatter is added to the halos before matching abundances with the stellar mass function, $\mgal$ does not enter in the parameterization at this stage. After adding scatter to $\vpeak$, the halos are re-rank-ordered and abundance matched as specified in Eq. (\ref{e.am}. A subsample of these models are shown in the figure. For each model, we measure $\slogmv$ as a function of $\vpeak$ from the mock galaxy distribution. As stated earlier, $\slogmv$ is not a constant in these models; it monotonically decreases with increasing $\vpeak$. Thus, this table allows us to look up the values of $\sigma[\mgal|\vpeak]$ as a function of $\vpeak$ required to make $\slogmv$ a constant for all $\vpeak$, as described here:

\begin{itemize}
\item{Select the value of $\slogmv$ for the model.}
\item{From the lookup table, determine $\sigma(\log  \vpeak)$ required at each $\vpeak$ to make $\sigma[\mgal|\vpeak]$ a constant for all halos. Figure \ref{scatter_tabulated_v1} shows reference lines for $\sigma[\mgal|\vpeak]=0.2$ and $0.4$.}
\item{For each halo, add a Gaussian random variable, $G(\sigma)$ to $\log\vpeak$. The width of the Gaussian is determined by the halo's $\vpeak$ value, from the previous step. }
\item{Rank-order the halos by this new quantity, $\log\vpeak+G(\sigma)$, and perform the abundance matching using Eq.~\eqref{e.am}. }
\end{itemize}

\bsp	
\label{lastpage}

\end{document}